\documentclass[12pt]{article}
\usepackage{epsfig,amsmath,amssymb,theorem,setspace}  
\newcommand{\zbyz}{\mathbb{Z}\times \mathbb{Z}}

\theoremstyle{plain}
\newtheorem{Rem}{Remark}

\begin{document}
\begin{titlepage}

\title{Homology of Fortuin-Kasteleyn clusters\\of Potts models on the torus}

\author{Louis-Pierre Arguin\thanks{E-mail:
    arguin@dms.umontreal.ca}\\D\'epartement de Physique, Universit\'e de Montr\'eal\\Case Postale 6128,
    succ. centre-ville\\Montr\'eal, Qu\'ebec\\Canada, H3C 1J7}
\maketitle

\end{titlepage}

\begin{abstract}
Topological properties of Fortuin-Kasteleyn clusters are studied on
the torus. Namely, the probability that their topology yields a given
subgroup of the first homology group of the torus is computed for
$Q=1$, $2$, $3$ and $4$. The expressions generalize those obtained by
Pinson for percolation ($Q=1$). Numerical results are also presented for
three tori of different moduli. They agree with the theoretical
predictions for $Q=1$, $2$ and $3$. For $Q=4$ agreement is not ruled
out but logarithmic corrections are probably present and they make it harder to decide.
\end{abstract}


\section{Introduction}

The Fortuin-Kasteleyn (FK) formulation of Potts models is often referred to
as the $Q$-state
bond-correlated percolation. The clusters formed by sites and bonds,
called FK clusters, may wrap the torus in a non-trivial way. One can
study the particular topology of a cluster by looking at its first
homology group. This group will be isomorphic to a subgroup of the
homology group of the
surface on which the model is defined. In general, consider $\mathcal{G}$, the topological space given by
all FK clusters of a configuration imbedded in a Riemann surface
$S$. There is a natural homomorphism $\phi : H_1(\mathcal{G})\rightarrow
H_1(S)$ between first homology group of each space. Therefore one may  look for the
probability that $\phi(H_1(\mathcal{G}))$ be a given subgroup $G$ of $H_1(S)$. A definition of this observable was first
given by Langlands {\it et al} in \cite{langlands}. In a preceding work, Pinson computed these
probabilities in the percolation case on the torus using a method
introduced by di Francesco {\it et al} \cite{diFran, pinson}. 

This paper generalizes Pinson's work on percolation to Potts models on
the torus for $Q=1$, $2$, $3$ and $4$ to compute the probability that a FK
configuration of a $Q$-state Potts model yields a particular
subgroup of the first homology group of the torus. This observable is
also studied numerically using Monte-Carlo simulations.

The general concepts required for our calculation are defined in section 2. Then essential points on
the transformation from the Potts model to a gaussian free field are
discussed before we deduce the explicit expressions for probabilities
in section 4. As expected, these are found to be modular
invariant. Numerical results presented for three tori with different
ratios are in good
agreement with values computed except for $Q=4$. For the latter case,
it is 
difficult to estimate the continuum limit numerically because what
looks like logarithmic finite-size corrections seem to appear.

\section{Definitions}

A FK configuration (or FK subgraph) $\mathcal{G}$ is built on a Potts configuration $\sigma$ by
putting bonds between all same-state neighbors and then removing them
with probability $1-p\equiv 1-\exp(-K)$ where $K$ is the critical
ratio $J/k_BT_c$. The resulting
graph $\mathcal{G}$ is said a FK subgraph associated to
$\sigma$. Neither correspondences $\sigma \rightarrow \mathcal{G}$ nor
$\mathcal{G} \rightarrow \sigma$ are unique. A maximally connected component
in $\mathcal{G}$ is called a FK cluster.

In this paper, we define the model on the torus,
i.e. $\mathbb{C}/(\mathbb{Z}+ \mathbb{Z}\tau$), $\tau\in \mathbb{Z}$. The first
homology group of the torus $T$ is isomorphic to $\zbyz$. Therefore its subgroups are isomorphic to $\{0\}$, $\zbyz$ and $\{m,n\} \equiv
\langle(m,n)\rangle$ where $m$, $n\in \mathbb{Z}$. 

Let $\alpha$ and $\beta$ be two independent cycles on the torus. On
the parallelogram $0$, $\tau$, $1$, $1+\tau$, we choose $\alpha$ to be
horizontal and $\beta$ to be parallel to the edge $(0,\tau)$. We distinguish different classes of FK clusters. They can be homotopic to a point. They
may wind the torus $a$ times along $\alpha$ and $b$ times along $\beta$, where $a$, $b\in \mathbb{Z}$ and $a\wedge b\equiv$gcd$(a,b)$ must be one. A
cluster of this type will be called a cluster $\{a,b\}$. We define
the signs of $a$ and $b$ as follows. Let us say we draw a boundary
curve of such a cluster. Then $a>0$ if we go from left to right and
$b>0$ if we go from top to bottom. Hence a
cluster $\{a,b\}$ is equivalent to a cluster $\{-a,-b\}$. The last
class is formed by cluster with a cross topology, precisely clusters
containing at least two connected subcomponents with different
non-trivial windings $\{a,b\}$.

\begin{figure}
\begin{center}
\leavevmode
\includegraphics[clip,width = 10cm]{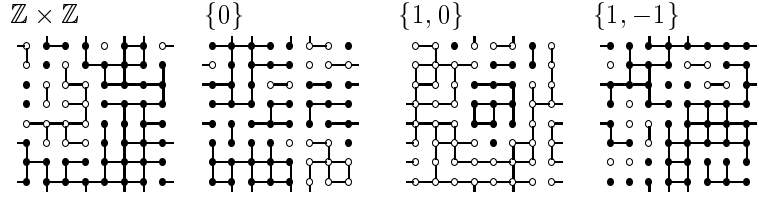}
\end{center}
\caption{Different $Q=2$ configurations with their associated subgroup. }
\label{fig:subgraphs}
\end{figure}

We consider the natural homomorphism $\phi : H_1(\mathcal{G})\rightarrow
H_1(T)$ for a FK subgraph
$\mathcal{G}$ and the torus $T$. We are interested in the subgroup $G$
of $\zbyz$ which lie in $\phi(H_1(\mathcal{G}))$. Then, we associate to $\mathcal{G}$
this particular subgroup $G$. In fact, the correspondence is:
\begin{itemize}

\item {$G$=$\{0\}$} if  $\mathcal{G}$ contains only clusters homotopic
  to a point;
\item {$G$=$\{a,b\}$} if $\mathcal{G}$ has only clusters homotopic to
  a point and
  at least one cluster of class $\{a,b\}$;
\item {$G$=$\zbyz$} if $\mathcal{G}$ contains a cross.
\end{itemize} 
Some examples are given in figure \ref{fig:subgraphs} for $Q=2$. The
  two states are depicted respectively by black dots and white
  dots. Bonds were added between same-state neighbors with
  probability $p$. The subgraphs associated with each subgroup will
  be  respectively
called: trivial subgraphs, subgraphs $\{a,b\}$ and subgraphs with a
cross topology. Because of the nature of FK clusters the probability of
$G=\{a,b\}$, $a\wedge b\neq 1$ is zero. Hence we will assume in the rest
of the paper that $a\wedge b=1$.

Let $Z_Q(G)$ be the partition function restricted to FK configurations
associated to the subgroup $G$ in the continuum limit. The probability that a subgraph yields $G$ is then 
\begin{equation}
\pi_Q(G)=\frac{Z_Q(G)}{Z_Q}
\label{pig}
\end{equation}
where $Z_Q$ is $Z_Q(\{0\})+Z_Q(\zbyz)+\sum_{a\wedge b=1}Z_Q(\{a,b\})$.

\section{Fortuin-Kasteleyn clusters and gaussian free fields}

To compute each weight, a tranformation of the Potts model in the
FK formulation to the {\sl F} model (or six-vertex
model) and finally to the {\sl SOS} model will be performed. The renormalization of the {\sl
  SOS} model into a bosonic free fields is then used to get the
probabilities in the continuum limit.

The first step of the transformation involves a decomposition of subgraphs into oriented-polygon decompositions. There is a bijection between subgraphs and possible non-oriented polygon decompositions on a lattice of $M\times N$ sites \cite{Baxter}. The partition function of the Potts model in the FK formulation can then be rewritten as a sum of separated weights at the critical temperature \cite{diFran,pinson}:
\begin{eqnarray}
\mathcal{Z} ^{M\times N}_Q      
&=&\mathcal{Z}^{M\times N}_Q(\{0\})+\mathcal{Z}^{M\times N}_Q(\zbyz)+\sum_{a\wedge b=1}\mathcal{Z}^{M\times N}_Q(\{a,b\})\nonumber\\
&=&\sum_{\mathcal{P}\{0\}}Q^{N_p/2}+Q\sum_{\mathcal{P}\{\mathbb{Z}\times\mathbb{Z}\}}Q^{N_p/2}+\sum_{a\wedge b=1}\sum_{\mathcal{P}\{a,b\}}Q^{N_p/2}.
\label{zp}
\end{eqnarray}
where $\mathcal{P}\{G\}$ stands for all decompositions whose subgraph
generates $G$. Here, $N_p$ stands for the number of polygons in a
specific decomposition. The $Q$ factor in the restricted weight of
$\zbyz$ comes from a different Euler relation between the number of sites,
polygons, connected components and faces for connected components with
a cross topology. 

One can build a bijective correspondence between trivial subgraphs and
subgraphs with a cross topology which preserves the number of polygons
in the polygon decomposition. Let $\mathcal{G}_+$ be a subgraph of the
cross topology type with decomposition $\mathcal{P}$. A trivial
subgraph $\mathcal{G}_0$ is built on the dual lattice of
$\mathcal{G}_+$ keeping the same $\mathcal{P}$ by putting its sites in
the middle of each smallest square formed by four sites of
$\mathcal{G}_+$. Then, on the dual lattice (sites in the middle of
plaquettes), a subgraph $\mathcal{G}_0$ is constructed by putting all
bonds that do not intersect with bonds of $\mathcal{G}_+$ on the
original lattice. The correspondence is bijective by construction and
$\mathcal{G}_0$ is necessarily trivial. One can also see that
$\mathcal{G}_+$ and $\mathcal{G}_0$ have the same number of polygons. Consequently, we obtain an important duality relation for weights of $\{0\}$ and $\zbyz$ in (\ref{zp}):
\begin{equation}
\mathcal{Z}^{M\times N}_Q (\zbyz)=Q\mathcal{Z} ^{M\times N}_Q (\{0\}).
\label{duality}
\end{equation}

In the {\sl F} model, we express the factor $Q^{{N_p}/2}$ in terms of a random
orientation on each polygon. Let us write $\epsilon_i=\pm 1$ for
non-trivial polygons (type {\sl n-h}) and $\epsilon_j=\pm 1$ for polygons homotopic to a point (type {\sl h}). We impose:
\begin{equation}
Q^{N_p/2}=Q^{N_{n-h}/2}Q^{N_h/2}=\sum_{\{\epsilon_i=\pm1\}}\cos\left[\frac{\pi}{2}e_0\sum_{i}\epsilon_i\right]\sum_{\{\epsilon_j=\pm1\}}\exp\left[4i\lambda\sum_j
  \epsilon_j\right]
\label{correction}
\end{equation}
where $e_0$ and $\lambda$ are chosen such as $Q^{1/2}=2\cos(\pi
e_0/2)$ and $Q=2\cos(4\lambda)$. We do not need the last details of the transformation to
the {\sl F} model for our purpose, but the reader could find them in \cite{Baxter}.

To construct the {\sl SOS} model from the {\sl F} one,  a field must
be defined on the faces of the lattice 
\cite{diFran,pinson}. These faces are formed by polygonal curves of
type {\sl n-h} or {\sl h}. A polygonal
line means an increase or a decrease by a factor $\pi/2$ of the field depending on the
polygon orientation. Discontinuities or frustrations are therefore
induced in the field by polygons {\sl n-h} of the decomposition. For
subgraphs with a cross topology and trivial one, only trivial polygons
are present; thus the field is always periodic. For subgraphs
$\{a,b\}$, if we write respectively $\delta\phi_1$ and
$\delta\phi_\tau$ for discontinuities observed along the cycles $\alpha$ and $\beta$, we have :
\begin{eqnarray}
\delta\phi_1=\frac{\pi}{2}b\sum_i\epsilon_i=\pi m & &\delta\phi_\tau=\frac{\pi}{2}a\sum_i\epsilon_i=\pi m'\label{delta}
\end{eqnarray}
where $m$,
$m'\in\mathbb{Z}$ and the sum is only over {\sl n-h} polygons. For subgraphs wih a cross
topology and trivial one, $m$ and $m'$ are zero. Subgraphs $\{a,b\}$
have possibly polygons {\sl h} and necessarily two polygons {\sl n-h} of class
$\pm[a,b]$ for each non-trivial connected component. Therefore, their
boundary conditions will be: $m=bk$, $m'=ak$ for a certain $k\in
\mathbb{Z}$. Moreover
\begin{Rem}
because $a\wedge b=1$, only a unique subgroup $\{a,b\}$ can produce non-periodic field with given discontinuities $m$, $m'$, $(m,m')\not=0$.
\label{remark}
\end{Rem}
We can express the correction term due to non-trivial polygons in
(\ref{correction}) as $\cos\left[\frac{\pi}{2}e_0\sum_i
  \epsilon_i\right]=\cos[\pi e_0 (m\wedge m')]$ with the help of
(\ref{delta}) and the fact that $a\wedge b=1$.

The extra factor $Q$ in front of the second term in (\ref{zp}) does not appear in the usual partition
function of the {\sl F} model. We have not taken it into account yet
in the {\sl SOS} model partition function. Consequently, let us first consider $\hat{\mathcal{Z}}^{M\times N}_Q$, the partition
function of the $Q$-state Potts model
with a weight $Q$ times lower than the correct one for subgraphs with a cross topology. In regard to the transformation performed, it corresponds to the {\sl SOS} model partition
function with the non-trivial correction. Considering that the {\sl SOS} model is assumed to renormalize onto a gaussian free
field, $\hat{\mathcal{Z}}^{M\times N}_Q$ becomes in the continuum limit
\cite{diFran}:

\begin{equation}
\hat{\mathcal{Z}}^{M\times N}_Q(e_0)\rightarrow \hat{Z}(g,e_0)\equiv\sum_{m,m'\in \mathbb{Z}}Z_{m,m'}(g/4)\cos\left[\pi e_0(m\wedge m')\right].
\label{zhat}
\end{equation}
The hat is the reminder of the missing factor $Q$. The value of $g$ is
given by $Q=2+2\cos\left(\pi g/2\right)$, $g\in[2,4]$ (see \cite{nienhuis}). The term $Z_{m,m'}(g)$ is the bosonic partition function on a torus $\tau=\tau_R+i\tau_I$ for a field with discontinuity $\delta_1\phi=2\pi m$, $\delta_\tau\phi=2\pi m'$ and coupling constant $g$:
\begin{equation}
Z_{m,m'}(g)=\frac{\sqrt{g}}{{\tau_I}^{1/2}
  |\eta(q)|^2}\exp\left[-\pi g\frac{m^2\tau_I^2 + (m'-m\tau_R)^2}{\tau_I}\right]
\label{zboson}
\end{equation}
where $q=\exp(2i\pi\tau)$ and $\eta(q)$ is the Dedekind function. 

\section{Probabilities of homology subgroups}

Each specific weight in (\ref{zp}) can be derived using expressions of
the last section. The partition function is then easily computed to
get the probability (\ref{pig}) for a particular subgroup $G$. Every expression found must be modular invariant.

We first find the weights of $\{0\}$ and $\zbyz$. We notice that for $e_0=1$ in (\ref{correction}), the quantity $Q^{N_{n-h}/2}$ is
zero , therefore $Z_Q(\{a,b\})\Big|_{e_0=1}=0$. Hence, if
we use this in (\ref{zhat}), we get:
\begin{equation}
\hat{Z}(g,1)=Z_Q(\{0\})+\frac{1}{Q}Z_Q(\zbyz).
\end{equation}
The desired results then follow from (\ref{duality})
\begin{eqnarray}
Z_Q(\{0\})&=&\frac{1}{2}\sum_{m,m'\in \mathbb{Z}}Z_{m,m'}(g/4)\cos\left[\pi (m\wedge m')\right]\label{0}\\
Z_Q(\zbyz)&=&\frac{Q}{2}\sum_{m,m'\in \mathbb{Z}}Z_{m,m'}(g/4)\cos\left[\pi (m\wedge m')\right]\label{+}.
\end{eqnarray}

We introduce the notation $Z_{0,0}^{a,b}(g/4)$ for the weight of all
subgraphs $\{a,b\}$ which induce periodic field and write
$\tilde{Z}_{m,m'}(g/4,e_0)\equiv Z_{m,m'}(g/4)\cos\left[\pi
  e_0(m\wedge m')\right]$ for convenience. The weight of the subgroup
$\{a,b\}$ is then given by the sum over all non-zero boundary conditions
particular to this subgroup by remark \ref{remark} plus the contribution of subgraphs $\{a,b\}$ with periodic field:
\begin{equation}
Z_Q(\{a,b\})=Z_{0,0}^{a,b}(g/4)+\sum_{\substack{m=bk\\m'=ak\\ k\in \mathbb{Z}\backslash\{0\}}}
                \tilde{Z}_{m,m'}(g/4,e_0).
\label{zab1}
\end{equation}
For $e_0=1$, it follows directly from (\ref{zab1}) that:
\begin{equation}
Z^{a,b}_{0,0}(g/4)=-\sum_{\substack{m=bk\\m'=ak\\k\in \mathbb{Z}\backslash\{0\}}}\tilde{Z}_{m,m'}(g/4,1).
\label{pi}
\end{equation}
We insert this new expression for $Z^{a,b}_{0,0}(g/4)$ in (\ref{zab1}). The correct weight is then given by:
\begin{equation}
Z_Q(\{a,b\})=\sum_{\substack{m=bk\\m'=ak\\k\in \mathbb{Z}\backslash\{0\}}}\left[\tilde{Z}_{m,m'}(g/4,e_0)-\tilde{Z}_{m,m'}(g/4,1)\right].
\label{zab2}
\end{equation}

The partition function is computed by summing over all subgroups:
\begin{eqnarray}
 Z_Q  &=&\frac{(Q+1)}{2}\hat{Z}(g,1)+\sum_{a\wedge b=1}\sum_{\substack{m=bk\\m'=ak\\k\in \mathbb{Z}\backslash\{0\}}}\left[\tilde{Z}_{m,m'}(g/4,e_0)-\tilde{Z}_{m,m'}(g/4,1)\right]\nonumber\\
        &=&\hat{Z}(g,e_0)+\frac{(Q-1)}{2}\hat{Z}[g,1]
\label{zq}
\end{eqnarray}
in agreement with reference \cite{diFran}. We added zero to the r.h.s of
the first equality in (\ref{zq}) by considering the term $m=m'=0$ in
the sum.

Each probability can then be calculated with (\ref{pig}). For a
specific $Q$, sums over $\tilde{Z}(g/4,e_0)$ involved in the
expressions found are computed using restricted sums over equivalence
class of $m\wedge m'$ so the cosinus is a constant in each of
them. 

The subgroups $\{0\}$ and $\zbyz$ are both stable under the action of
any modular transformation of the torus whose generators are
$T:\tau\rightarrow\tau+1$ and $S:\tau\rightarrow -1/\tau$. Hence,
their probability must be invariant. For subgraphs $\{a,b\}$, modular
transformations of non-trivial curves impose:
\begin{eqnarray}
\pi_Q(\{(a,b)\})\Big|_\tau&=&\pi_Q(\{(a+b,b)\})\Big|_{\tau+1}\\
\pi_Q(\{(a,b)\})\Big|_\tau&=&\pi_Q(\{(-b,a)\})\Big|_{-1/\tau}.
\end{eqnarray}
These restrictions are easily verified directly on (\ref{zab2}) with the help of the modular transformation of the bosonic partition function \cite{CFT}:
\begin{eqnarray}
Z_{m,m'}\Big|_\tau&=&Z_{m,m+m'}\Big|_{\tau+1}\\
Z_{m,m'}\Big|_\tau&=&Z_{m',-m}\Big|_{-1/\tau}.
\end{eqnarray}

In the particular cases $Q=1$ and $Q=2$, the expressions for
$\pi_Q(\{a,b\})$ are written in an elegant way using theta functions
\cite{CFT}. The calculations yield for $\tau_{a,b}\equiv a-b\tau$
\begin{eqnarray}
\pi_1(\{a,b\})&=&\frac{1}{2|\tau_{a,b}||\eta(q)|^2}\left[\theta_3\left(\frac{i\tau_I}{6|\tau_{a,b}|^2}\right)-\theta_3\left(\frac{3i\tau_I}{2|\tau_{a,b}|^2}\right)-2\theta_2\left(\frac{3i\tau_I}{2|\tau_{a,b}|^2}\right)\right]\\\nonumber
 \\ 
\pi_2(\{a,b\})&=&\frac{1}{|\tau_{a,b}||\eta(q)|}\left[\frac{\theta_2\left(\frac{i\tau_I}{3|\tau_{a,b}|^2}\right)-\theta_3\left(\frac{i\tau_I}{3|\tau_{a,b}|^2}\right)+\theta_4\left(\frac{i\tau_I}{3|\tau_{a,b}|^2}\right)}{|\theta_2(\tau)|+|\theta_3(\tau)|+|\theta_4(\tau)|}\right].
\end{eqnarray}

\section{Numerical simulations} 

Numerical simulations were performed on square lattices with $V\times
H$ sites on three different tori, $\tau=i$, $\tau=2i$ and
$\tau=i+1/2$. We chose to consider the quantity
$\pi_Q(\{0\})+\pi_Q(\zbyz)$ instead of two different values because
they are simply related by (\ref{duality}) for every lattice. This
allowed us technically to experiment on larger samples. Results are
shown in Table \ref{tab:prob}. 

On a $V\times H$ lattice, the identification rules on the boundaries
for a site $(i,j)$ were $(i,H)\equiv(i,0)$, $(V,j)\equiv(0,j)$ for
$\tau=i$ and $\tau=2i$ and $(i,H)\equiv(i,0)$, $(V,j)=(0,j+V/2 \mod V)$ for
$\tau=i+1/2$. We used Swendsen-Wang algorithm for $Q=2$, $3$ and
$4$ with at least 8000 initial thermalisations and five upgrades
between measurements. 

We proceed differently to approximate
the continuum limit for $Q=1$, $2$ than for $Q=3$, $4$. In the first case, we took lattices of dimensions $256\times 256$
sites for $\tau=i$ and $\tau=i+1/2$,  $256\times 128$ for
$\tau=2i$. These are indeed good approximations of the continuum
limit as can be seen by comparing measurements on lattices with half
this linear size. In Table 1, the numbers after the vertical bar represent the statistical
error on the last two digits for a 95\%-confidence interval. For example, $0.0210|10$ means the
interval $[0.0200,0.220]$. Each probability was calculated on a
sample larger than $2\times 10^7$ configurations. The figures in bold
faces are the predictions obtained from (\ref{0}), (\ref{+}) and
(\ref{zab2}). The agreement
between numerical data and analytic calculations are convincing.

\begin{figure}
\begin{center}
\leavevmode
\includegraphics[clip,width = 12cm]{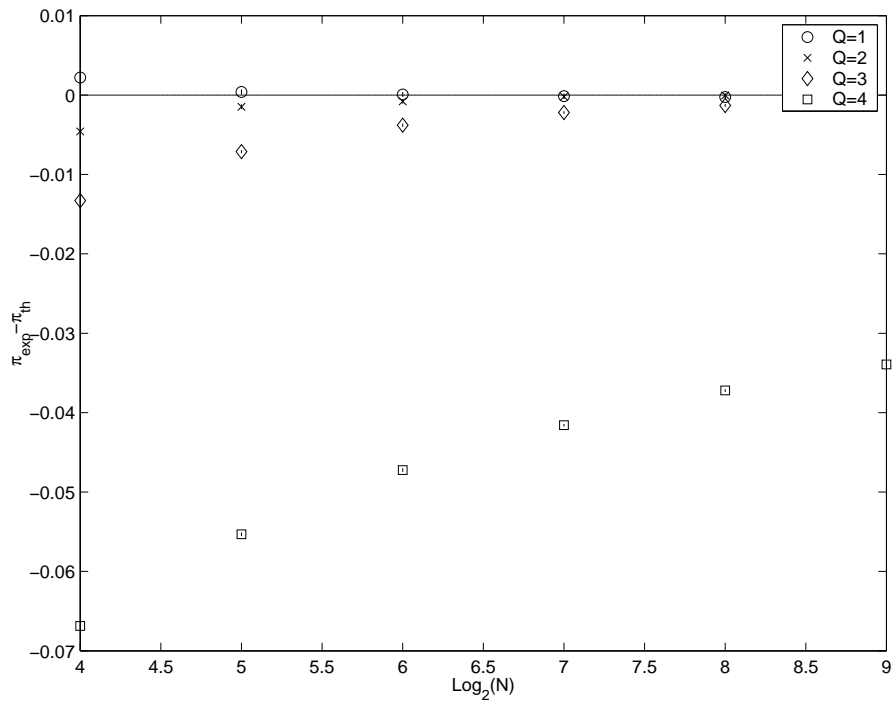}
\end{center}
\caption{Deviation of numerical results from theoretical prediction for  $\pi(\zbyz)+\pi({0})$ on a $N \times N$ lattice ($\tau = i$) as a function of $\log_2(N)$}
\label{deviation}
\end{figure}

For $Q=3$ and $Q=4$, a more complex method is necessary because the scale effect is still visible in large lattices. Consequently, we used a power law hypothesis \cite{saint}:
\begin{equation}
|\pi_Q(G,H,V)-\pi_Q(G)|\sim \alpha V^{\beta}.
\label{approx}
\end{equation}
where $\alpha>0$ and $\beta<0$ are two parameters
to be determined. We obtained $\alpha$ and $\beta$ by a linear
regression of the function $f[i]\equiv
\log|\pi_Q({a,b},2^iH_0,2^iV_0)-\pi_Q(\{(a,b)\},2^{i+1}H_0,2^{i+1}V_0)|$.
We took $V_0=16$ for $Q=3$ and $V_0=32$ for $Q=4$
with $i=0,1,2,3,4$.
The errors were estimated using a numerical method. For each $i$, we
had a sample of at least $2\times 10^7$ configurations. The results are very good for $Q=3$ although errors are larger
using this method. One might ask whether the exponent $\beta $ of
(\ref{approx}) is
universal. However, the values of the exponent $\beta$, say in the case
$Q=3$, differed for the different subgroups and the different ratios
studied though they remained of order $1$. For $Q=4$, we naively tried the same
approximation. These results in Table 1 and the behaviour
plotted in figure \ref{deviation} suggest that the finite-size
correction term is not of the form (\ref{approx}). Several authors
have argued convincingly that finite-size effects for the $Q=4$ Potts
model show logarithmic corrections \cite{Kleban}. It is hard to see them in our data
since the fit would involve more parameters and thus would need more points to
be accurate. Thus one would have to
experiment on lattices larger than $512\times 512$ and to deal with
the problem of getting samples with a relatively small statistical error.

\section*{Acknowlegments}
I would like to thank Yvan Saint-Aubin, first because he gave
me this
problem, and second for his very helpful advice. This work was made
possible by a NSERC fellowship.

\newcommand{\bo}{\bfseries}


\pagebreak
\singlespacing
\begin{table}[t]
\caption{Probabilities calculated with both {\bo analytical} and numerical method. }
\begin{center}
\leavevmode

\begin{tabular}{lllllll}
\hline\hline
$Q$&$\tau$&$\pi_Q(\{0\})+\pi_Q(\zbyz)$&$\pi_Q(\{1,0\})$&$\pi_Q(\{0,1\})$&$\pi_Q(\{1,1\})$&$\pi_Q(\{1,-1\})$\\
\hline

{\large 1}&$i$&{\bo .61908}&{\bo .16942}&{\bo .16942}&{\bo .020980}& {\bo .020980}\\
&&.61903$|$22&.16948$|$17&.16939$|$17&.020995$|$64&.020957$|$ 64\\
&$2i$&{\bo.45376}&{\bo.50304}&{\bo.024950}&{\bo.008755}&{\bo.008755}\\
&&.45380$|$22&.50293$|$22&.024983$|$70&.008749$|$42&.008779$|$42\\
&$i+1/2$&{\bo.62840}&{\bo.16816}&{\bo.10006}&{\bo.10006}&{\bo.001519}\\
&&.62849$|$22&.16812$|$17&.10008$|$13&.09998$|$13&.001504$|$17\\
\hline
{\large 2}&$i$&{\bo.67927}&{\bo.14644}&{\bo.14644}&{\bo.013903}&{\bo.013903}\\
&&.67907$|$21&.14654$|$16&.14645$|$16&.013949$|$53&.013940$|$53\\
&$2i$&{\bo.47863}&{\bo.49625}&{\bo.015376}&{\bo.004734}&{\bo.004734}\\
&&.47888$|$32&.49595$|$32&.015384$|$78&.004751$|$43&.004760$|$43\\
&$i+1/2$&{\bo.69083}&{\bo.14569}&{\bo.08095}&{\bo.08095}&{\bo.000727}\\
&&.69091$|$22&.14569$|$17&.08100$|$13&.08081$|$13&.000724$|$13\\
\hline
{\large 3}&$i$&{\bo.74533}&{\bo.11867}&{\bo.11867}&{\bo.008660}&{\bo.008660}\\
&&.7457$|$21&.1173$|$19&.1187$|$14&.00863$|$42&.00871$|$49\\
&$2i$&{\bo.52571}&{\bo.45995}&{\bo.0092445}&{\bo.0024968}&{\bo.0024968}\\
&&.5250$|$21&.4605$|$21&.00921$|$45&.00240$|$29&.00247$|$23\\
&$i+1/2$&{\bo.75821}&{\bo.11820}&{\bo.061448}&{\bo.061448}&{\bo.00033}\\
&&.7584$|$19&.1180$|$15&.0606$|$13&.0608$|$13&.00000$|$81\\
\hline
{\large 4}&$i$&{\bo.85034}&{\bo.071728}&{\bo.071728}&{\bo.0030997}&{\bo.0030997}\\
&&.8265$|$23&.08343$|$16&.08151$|$16&.00380$|$43&.00440$|$32\\
&$2i$&{\bo.63368}&{\bo.36172}&{\bo.0032494}&{\bo.00067549}&{\bo.00067549}\\
&&.6013$|$25&.3929$|$25&.00369$|$28&.00087$|$16&.00085$|$14\\
&$i+1/2$&{\bo.86316}&{\bo.07151}&{\bo.03260}&{\bo.03260}&{\bo.000061}\\
&&.8426$|$23&.0794$|$19&.0389$|$11&.0383$|$13&.00011$|$4\\
\hline\hline
\end{tabular}

\label{tab:prob}
\end{center}
\end{table}

\end{document}